\documentclass[conference]{IEEEtran}
\IEEEoverridecommandlockouts
\usepackage{cite}
\usepackage{amsmath,amssymb,amsfonts}
\usepackage{algorithmic}
\usepackage{graphicx}
\usepackage{textcomp}
\usepackage{booktabs}
\usepackage{comment}

\usepackage{xcolor}
\def\BibTeX{{\rm B\kern-.05em{\sc i\kern-.025em b}\kern-.08em
    T\kern-.1667em\lower.7ex\hbox{E}\kern-.125emX}}
    
\usepackage{xcolor}
\usepackage{algorithm,letltxmacro}
\newcommand{\Order}[1]{\ensuremath{\mathcal{O}(#1)}}    

\begin{document}

\title{Landau collision operator in the CUDA programming model applied to thermal quench plasmas}

\author{\IEEEauthorblockN{1\textsuperscript{st} Mark F. Adams}
\IEEEauthorblockA{\textit{Lawrence Berkeley National Laboratory} \\
mfadams@lbl.gov}
\and
\IEEEauthorblockN{2\textsuperscript{nd} Dylan P. Brennan}
\IEEEauthorblockA{\textit{Princeton University} \\
dylanb@princeton.edu}
\and
\IEEEauthorblockN{3\textsuperscript{rd} Matthew G. Knepley}
\IEEEauthorblockA{\textit{University at Buffalo} \\
knepley@gmail.com}
\and
\IEEEauthorblockN{4\textsuperscript{th} Peng Wang}
\IEEEauthorblockA{\textit{NVIDIA} Corporation \\
penwang@nvidia.com}
}

\maketitle

\begin{abstract}
Collisional processes are critical in the understanding of non-Maxwellian plasmas.
The Landau form of the Fokker-Planck equation is the gold standard for modeling collisions in most plasmas, however $\Order{N^2}$ work complexity inhibits its widespread use.
We show that with advanced numerical methods and GPU hardware this cost can be effectively mitigated.
This paper extends previous work on a conservative, high order accurate, finite element discretization with adaptive mesh refinement of the Landau operator, with extensions to GPU hardware and implementations in both the CUDA and Kokkos programming languages.
This work focuses on the Landau kernels and on NVIDIA hardware, however preliminary results on AMD and Fujitsu/ARM hardware, as well as end-to-end performance of a velocity space model of a plasma thermal quench, are also presented.
Both the fully implicit Landau time integrator and the plasma thermal quench model are publicly available in PETSc (Portable, Extensible, Toolkit for Scientific computing).
\end{abstract}


\section{Introduction}

The Vlasov-Maxwell-Boltzmann system of equations is the fundamental model of magnetized plasmas.
It evolves a distribution function for each species (one electron and potentially many ions species) in phase space with up to three configuration space dimensions plus three velocity space dimensions.
The Fokker-Planck (FP) equation is a computationally tractable expansion of the Boltzmann equation \cite{Kramers1940BrownianMI,Moyal1949StochasticPA} that includes only grazing Coulomb collisions, which is effective when collisional effects are dominated by small angle deviations, as is common in most plasmas.
The Landau form of FP conserves density, momentum and energy and admits unstructured finite element discretizations that conserve these quantities exactly \cite{landau1936kinetic,Hirvijoki2016}, however it is an $\Order{N^2}$ work complexity algorithm.
Alternatively, a Rosenbluth potentials formulation of FP is asymptotically less expensive with an optimal solver, with two Laplacian solves per species per nonlinear iteration, but conserves energy only asymptotically \cite{Rosenbluth1957,Chacn2000AnIE,Shiroto2020AMD}.
This paper builds on previous work, that used vector processing \cite{AdamsHirvijokiKnepleyBrownIsaacMills2017}, with the use of GPUs and shows that the Landau algorithm can be practical, given that velocity space meshes are inherently not large, especially with mesh adaptivity and high order accurate discretizations.

Accurate FP collisions are critical in modeling many important processes in plasmas, such as the generation of highly structured non-Maxwellian distributions during dynamical processes, and methods that conserve energy with arbitrary accuracy are critical for long time simulations.  One of the most important dynamical processes to study is the rapid cooling of the bulk of the distribution, a thermal quench.  If the quench occurs fast enough, the less collisional high energy tail of the original distribution will not cool as fast, and can form a high energy ``bump'' population on the tail of the distribution.  This type of distribution can lead to kinetic instabilities, and given a high enough electric field, can be accelerated to runaway conditions. 
A runaway electron event can cripple a fusion reactor for months and thereby threaten the mission of reactor scale experiments like ITER and the commercial viability of fusion power.

Contemporary high performance hardware for scientific computing falls into two broad categories: massively parallel GPUs coupled with CPUs and manycore vector processors, each coupled with distributed memory processing.
GPUs are characterized by hierarchical collaborative thread groups with hierarchical shared memory.
This architectural complexity requires new programming models and languages.
CUDA became the dominant language to support GPUs and its programming model is now supported by, for instance, HIP and SYCL, as well as Kokkos.
All of these languages implement the CUDA programming model, however Kokkos also generates code for manycore vector processors by mapping its {\it league members} to OpenMP threads, instead of CUDA {\it blocks}, by mapping its {\it thread team member's vector} threads to vector lanes instead of a CUDA {\it thread} dimension, and using only two levels of hierarchical parallelism. 
Kokkos thereby provides a portable programming language for the primary classes of today's high performance computing hardware.

This paper proceeds with a derivation of the Landau operator and a Vlasov-Poisson-Landau thermal quench model in \S\ref{sec:back}.
\S\ref{sec:numr} describes the numerical methods and software used in this work.
\S\ref{sec:physics} presents the physics motivation and demonstrates that our model generates the expected plasma dynamics.
\S\ref{sec:perf} examines the throughput performance of the plasma quench model on an IBM/NVIDIA V100 node, with CUDA and Kokkos-CUDA, an AMD EPYC/MI100 node (Kokkos-HIP), and a Fujitsu A64FX node (Kokkos-OpenMP), and hardware utilization on the V100.
\S\ref{sec:conc} concludes the report.

\section{Background}
\label{sec:back}
The evolution of the phase space distribution or density function $f\left( \vec{x}, \vec{v}, t\right)$ of a plasma in an electromagnetic field is effectively modeled with a Vlasov-Maxwell-Boltzmann system of the form
\begin{equation*}
\begin{split}
\frac{df}{dt} & \equiv
\frac{\partial f}{\partial t} + \frac{\partial  \vec{x}}{\partial t}
\cdot \nabla_x f+ \frac{\partial  \vec{v}}{\partial t} \cdot \nabla_v f \\
& = \frac{\partial f}{\partial t} + {  \vec{v}} \cdot \nabla_x f+
\frac{e}{m}\left( { \vec{E}} + {  \vec{v}} \times { \vec{B}} \right) \cdot
\nabla_v f = C
\end{split}
\end{equation*}
with charge $e$, mass $m$, electric field ${ \vec{E}} $, magnetic field ${ \vec{B}}$, spatial coordinate ${  \vec{x}}$ , velocity
coordinate $\vec{v}$  and a collision term $C$ \cite{Vlasov1968}.
This equation is composed of the symplectic {\it Vlasov-Maxwell} system $\frac{df}{dt}=0$ and a metric, or diffusive, collision operator $C$. 
For this presentation, assume ${B=0}$ and ignore configuration space.
The collision operator is in velocity space only.
A source term is added for the plasma quench model, and collisions are expanded for multiple species, resulting in, after dropping the gradient subscripts, species $\alpha$ evolving according to
\begin{equation}
\label{eq:landau1a}
\frac{\partial f_{\alpha}}{\partial t} + \frac{e_{\alpha}}{m_{\alpha}} { \vec{E}} \cdot \nabla f_{\alpha} = \sum_{\beta} C_{\alpha\beta} + S_\alpha\left(t\right).
\end{equation}
The Landau form of Fokker-Planck collisions for species $\alpha$, colliding with species $\beta$, is given by
\begin{equation}
\label{eq:landau1}
C_{\alpha\beta} = 
\nu_{\alpha\beta}\frac{m_0}{m_{\alpha}}\nabla \cdot\int \limits_{\bar\Omega} d{\bar{v}}\;\mathbf{U}(\vec{v},{\bar{v}})\cdot\left(\frac{m_0}{m_{\alpha}}\bar{f}_{\beta}\nabla f_{\alpha} - \frac{m_0}{m_{\beta}}f_{\alpha} \bar \nabla \bar{f}_{\beta}\right)
\end{equation}
with a collision frequency $\nu_{\alpha\beta}=e_{\alpha}^2e_{\beta}^2\ln\Lambda_{\alpha\beta}/8\pi m_0^2\varepsilon_0^2$, the Coulomb logarithm $\ln\Lambda_{\alpha\beta}$ (=10 herein), an arbitrary reference mass $m_0$ , the vacuum permittivity $\varepsilon_0$ and the effective charges $e$ of each species. 
Overbar terms are evaluated on the grid for the domain $\bar\Omega$ of species $\beta$ and $\bar v \equiv \vec{\bar{v}}$ for clarity. 
The Landau tensor $\mathbf{U}(\vec{v},{\bar{v}})$ is a scaled projection
matrix defined as
\begin{equation}
\label{eq:landau_tensor}
\mathbf{U}(\vec{v},{\bar{v}})=\frac{1}{\lvert
\vec{v}-{\bar{v}}\rvert^3}\left(\lvert \vec{v}-{\bar{v}}\rvert^2\mathbf{I}-(\vec{v}-{\bar{v}})(\vec{v}-{\bar{v}})\right).
\end{equation}
This system is nondimensionalized according to Appendix \ref{sec:nondim}.

\subsection{Weak form}

In this work, equation (\ref{eq:landau1a}) is written in cylindrical coordinates, $\vec{v} = \left ( r, z \right)$, where the electric field is aligned with the $z$ coordinate.
A full $3D$ model is supported in the library and is required for extension to relativistic regimes \cite{Beliaev_Budker_1956,Braams-Karney:PRL1987}.
The weak form of the evolution equation for species $\alpha$, given a test function $\psi({ \vec{v}})$ as derived in \cite{Hirvijoki2016}, can be expressed as
\begin{equation}
\label{eq:weak-form}
\begin{aligned}
2\pi \int \limits_{\Omega} d{ \vec{v}} r \psi \cdot \left ( \frac{\partial  f_{\alpha}}{\partial t}  + \left ( 0,  \frac{ e_\alpha}{m_{\alpha}} E_z \right)  \cdot \nabla f_{\alpha}  \right)  = \\ \sum_{\beta}\left(\psi,f_{\alpha}\right)_{\mathbf{D},\alpha\beta} + \sum_{\beta}\left(\psi, f_{\alpha}\right)_{\mathbf{K},\alpha\beta}+\left(\psi,S_\alpha\right) ,
\end{aligned}
\end{equation}
where $(\cdot,\cdot)_{\Omega}$ is the $L^2$ inner product in $\Omega$ and ${\vec{E} = E_{z}} \hat{z}$.
Using integration by parts the inner products of the two parts of the Landau collision integral for species $\alpha$ can be expressed as
\begin{align}
\label{eq:D}
\left(\psi,\phi\right)_{\mathbf{D},\alpha\beta}&=-\int \limits_{\Omega}d{ \vec{v}}r\nabla\psi\cdot{\nu}_{\alpha\beta}\frac{m_0}{m_{\alpha}}\frac{m_0}{m_{\alpha}}\mathbf{D}(f_{\beta},{ \vec{v}})\cdot\nabla\phi \\
\label{eq:K}
\left(\psi, \phi\right)_{\mathbf{K},\alpha\beta}&=\int \limits_{\Omega}d{ \vec{v}}r\nabla\psi\cdot {\nu}_{\alpha\beta}\frac{m_0}{m_{\alpha}} \frac{m_0}{m_{\beta}}\mathbf{K}(f_{\beta},{ \vec{v}})\, \phi.
\end{align}
The tensor $\mathbf{D}$ and the vector $\mathbf{K}$ are defined as
\begin{align}
\label{eq:diff_coef}
  \mathbf{D}(f,{ \vec{v}})& \equiv \int
  \limits_{\bar\Omega}d{\bar{v}}\bar r\;\mathbf{U}^D({ \vec{v}},{\bar{v}})f({\bar{v}}), \\
\label{eq:fric_coef}
  \mathbf{K}(f,{ \vec{v}})& \equiv \int \limits_{\bar\Omega}d{\bar{v}}\bar r\;\mathbf{U}^K({ \vec{v}},{\bar{v}})\cdot\bar{\nabla}f({\bar{v}}), 
\end{align}
where $\mathbf{U}^D$ and $\mathbf{U}^K$ are forms of the Landau tensor in cylindrical coordinates.
These tensors are much more complex than (\ref{eq:landau_tensor}) \cite{Hirvijoki2016}. 

\section{Numerical methods}
\label{sec:numr}

The Vlasov-Maxwell-Landau system can be discretized with grid methods (Eulerian) or with particles (Lagrangian).
Particles are generally more efficient than grids for high dimensional problems.
For instance, with a second order accurate grid method and an \Order{N^{\frac{1}{2}}} accurate particle method, the complexity of the grid and particle methods cross-over at $4D$: halving the mesh spacing reduces the error by $4x$ and requires $2^4=16$ times more grid cells in $4D$, and $16$ times more particles reduces the error by $4x$ in any dimension.
The Landau operator presented here is entirely on a velocity space grid, however it can be use in a particle method with conservative particle-grid interpolation \cite{MollenJPP2021}.


Implicit time integrators are useful in the advance of the collision term, which requires a nonlinear solver.
The full linearization of the Landau operator is a dense matrix that would be prohibitively expensive to build and solve.
A practical approximate linearization is to compute $\mathbf{D}(f,\vec{v})$ and $\mathbf{K}(f,\vec{v})$ about the current state and applying standard finite element methods to (\ref{eq:D}) and (\ref{eq:K}).
A traditional Newton iteration is used with this approximate Jacobian, which is fully recomputed in each iteration.
This quasi-Newton iteration converges linearly, is robust and similar to the solver used in production in the XGC code \cite{Hager2016}.
This matrix has the property, unusual for a multiple degree-of-freedom Jacobian, that the species are not coupled.
With $S$ species and a single species Jacobian $A_1$, the non-zero pattern of $A_S$ is $I_{S \times S} \otimes A_1$.
Thus, the multi-species Landau Jacobian matrix is block diagonal.



In the remainder of this section, \S\ref{sec:loops} describes the transformation of the natural implementation of Landau to an optimal form, \S\ref{sec:amr} describes the mesh adaptivity methodology, the code structure is described in  \S\ref{sec:structure}, and the CUDA and Kokkos implementations are discussed in \S\ref{sec:cudakok}.

\subsection{Loop optimizations and CUDA algorithm}
\label{sec:loops}

To simplify the derivation of the optimal loop organization for the Landau kernel only the $\mathbf{K}_{\alpha\beta}$ term in the right hand side of (\ref{eq:weak-form}) is derived in detail.
The $\mathbf{D}_{\alpha\beta}$ term is treated similarly.

Start by factoring $\nu_{\alpha\beta}$, as $\nu_{\alpha\beta} = \nu e_\alpha^2 e^2_\beta$, and bring the sum over $\beta$ in (\ref{eq:weak-form}) into (\ref{eq:K}), to form:

\begin{equation}
\label{eq:KK}
  \sum_{\beta}\left(\psi, f_{\alpha}\right)_{\mathbf{K},\alpha\beta} = \int \limits_{\Omega}d\vec{v}r\nabla\psi\cdot \nu e_\alpha^2\frac{m_0}{m_\alpha} \sum_{\beta} e^2_\beta \frac{m_0}{m_{\beta}}\mathbf{K}(f_{\beta},\vec{v}) \, f_{\alpha}.
\end{equation}

Next, move the $\beta$ loop into the inner integral in the $\mathbf{K}(f_{\beta},\vec{v})$ term to form: 

\begin{equation}
\label{eq:KKK}
 \sum_{\beta} e^2_\beta \frac{m_0}{m_{\beta}}\mathbf{K}(f_{\beta},\vec{v})  = \int \limits_{\bar\Omega}d{\bar{v}} {\bar r}\;\mathbf{U}^K(\vec{v},{\bar{v}}) \cdot \sum_{\beta}  e^2_\beta \frac{m_0}{m_{\beta}} \bar{\nabla} f_\beta({\bar{v}}).
\end{equation}
From (\ref{eq:KK}) and (\ref{eq:KKK}), the $\mathbf{K_{\alpha\beta}}$ term  in (\ref{eq:weak-form}) is expressed as:
\begin{equation}
\label{eq:Kalpha}
\begin{aligned}
\sum_{\beta}& \left(\psi, f_{\alpha} \right)_{\mathbf{K},\alpha\beta} = \int \limits_{\Omega}d\vec{v}r\nabla\psi\cdot \\ 
 & \left [ \nu e_\alpha^2\frac{m_0}{m_\alpha}  \int \limits_{\bar\Omega}d{\bar{v}} {\bar r}\;\mathbf{U}^K(\vec{v},{\bar{v}}) \cdot  \sum_{\beta}  e^2_\beta \frac{m_0}{m_{\beta}} \bar{\nabla} f_\beta({\bar{v}}) \right ] \, f_{\alpha}.
\end{aligned}
\end{equation}
Equation (\ref{eq:Kalpha}) a standard finite element weak form and only the coefficient vector term in the bracket is unique to this operator.
Applying this processes to (\ref{eq:D}) results in a standard finite element discretization of the Laplacian with a coefficient tensor that is unique to this operator.


Algorithm \ref{algo:1} is CUDA pseudo code for the Landau Jacobian matrix construction for one element $e$ on one CUDA SM (or one league member in the Kokkos version) with $S$ species, $N_q$ integration points per element and $N$ global integration points.
The Kokkos version is similar (\S\ref{sec:cudakok}).

Arrays of coordinates $r$ and $z$, weights $w$, function values $f$, and gradients $df$ for each integration point are computed on the GPU to allow for efficient processing in the inner integral.
The element Jacobian $\mathbf{J}$ for the given element and finite element tablatures for the order of the element $\mathbf{B}$ and $\mathbf{E}$ are also provided.
\begin{algorithm}[h!]
\begin{algorithmic}[1]
\STATE{$i \leftarrow  threadIdx.y$} \COMMENT{local integration point index}
\STATE{$gi \leftarrow  e*N_q + i$} \COMMENT{global integration point}
\FOR[Integral over all integration points]{$j= threadIdx.x : blockDim.x:N$}
\STATE{$\left[\mathbf{U_K}, \mathbf{U_D} \right]  \leftarrow \mathbf{LandauTensor2D}\left(r[gi],z[gi],r[j],z[j]\right)$  }
\FOR{$\beta=1:S$}
\STATE{$\mathbf{T_K}  \leftarrow \mathbf{T_K} + e^2_\beta \frac{m_o}{m_{\beta}} df[:][\beta][j]$}
\STATE{${T_D}  \leftarrow {T_D} + e^2_\beta f[\beta][j]$}
\ENDFOR
\STATE{$\mathbf{G_K} \leftarrow \mathbf{G_K} + w[j] \mathbf{U_K} \cdot \mathbf{T_K} $}
\STATE{$\mathbf{G_D}  \leftarrow \mathbf{G_D} + w[j] {T_D} \mathbf{U_D} $}
\ENDFOR
\STATE{Reduce $\mathbf{G_K}$ and  $\mathbf{G_D}$ across threads}
\FOR{$\alpha=threadIdx.x:blockDim.x:S$}
\STATE{$\mathbf{K_i}\left[{\alpha}\right]  \leftarrow {\nu} e^2_\alpha \frac{m_o}{m_{\alpha}}\mathbf{G_K} $}
\STATE{$\mathbf{D_i}\left[{\alpha}\right]  \leftarrow - {\nu} e^2_\alpha \left( \frac{m_o}{m_{\alpha}} \right)^2\mathbf{G_D}$}
\ENDFOR
\STATE{syncthreads}
\FOR{$\alpha=threadIdx.x:blockDim.x:S$}
\STATE{$\mathbf{KK}\left[{\alpha}\right]\left[i\right] \leftarrow \mathbf{J}\left(q_i\right)^{-1} \mathbf{K_i}\left[{\alpha}\right]  w[gi]$} \COMMENT{to global basis}
\STATE{$\mathbf{DD}\left[{\alpha}\right]\left[i\right]  \leftarrow \mathbf{J}\left(q_i\right)^{-1} \mathbf{D_i}\left[{\alpha}\right]  \mathbf{J}\left(q_i\right)^{-1} w[gi]$}
\ENDFOR
\STATE{syncthreads}  \COMMENT{use all threads to assemble element matrix}
\STATE{$\mathbf{C} \leftarrow Transform\&Assemble\left({0}, \mathbf{KK}, \mathbf{DD}, \mathbf{B}, \mathbf{E} \right)$}
\end{algorithmic} 
\caption{Build one element Jacobian matrix $\mathbf{C}$ on one SM with CUDA syntax}
\label{algo:1}
\end{algorithm}
The assembly of element matrix $\mathbf{C}$ into the global matrix, with interpolation of constrained vertices to unconstrained vertices that result from the mesh adaptivity method, is not shown.

Critically, this formulation removes $\alpha$ terms from the inner integration loop, which allows for a loop over one species in the leading complexity term, resulting in a complexity of \Order{N_e N N_q S}, where $N_e$ is the number of elements ($N \equiv N_e N_q$), or simply \Order{N^2 S}. 
The complexity of $Transform\&Assemble$ is \Order{N_e N_b^2 N_q S}, or \Order{N N_b^2 S}, where $N_b$ is the number of vertices per element, which is equal to $N_q$ for the tensor elements used herein (e.g., $N_q=16$).

\subsection{Adaptive mesh refinement}
\label{sec:amr}

The $\Order{N^2}$ complexity of Landau can be mitigated by first adapting the grid to place points so as to represent the solution most efficiently as presented in \cite{AdamsHirvijokiKnepleyBrownIsaacMills2017}.
The {\it p4est} library is used in this work \cite{Stadler1033,DBLP:journals/siamsc/IsaacBWG15,Rudi:2015:EIS:2807591.2807675}.
{The Landau solver provides a high-level parameterization of mesh adaptivity, with command line options, to generate grids for Maxwellian distributions} and for common runaway electron distributions.
Figure \ref{fig:grids} shows a typical mesh of a two species plasma with Maxwellian distributions.
\begin{figure}[htbp]
\begin{center}
\includegraphics[width=1.\linewidth]{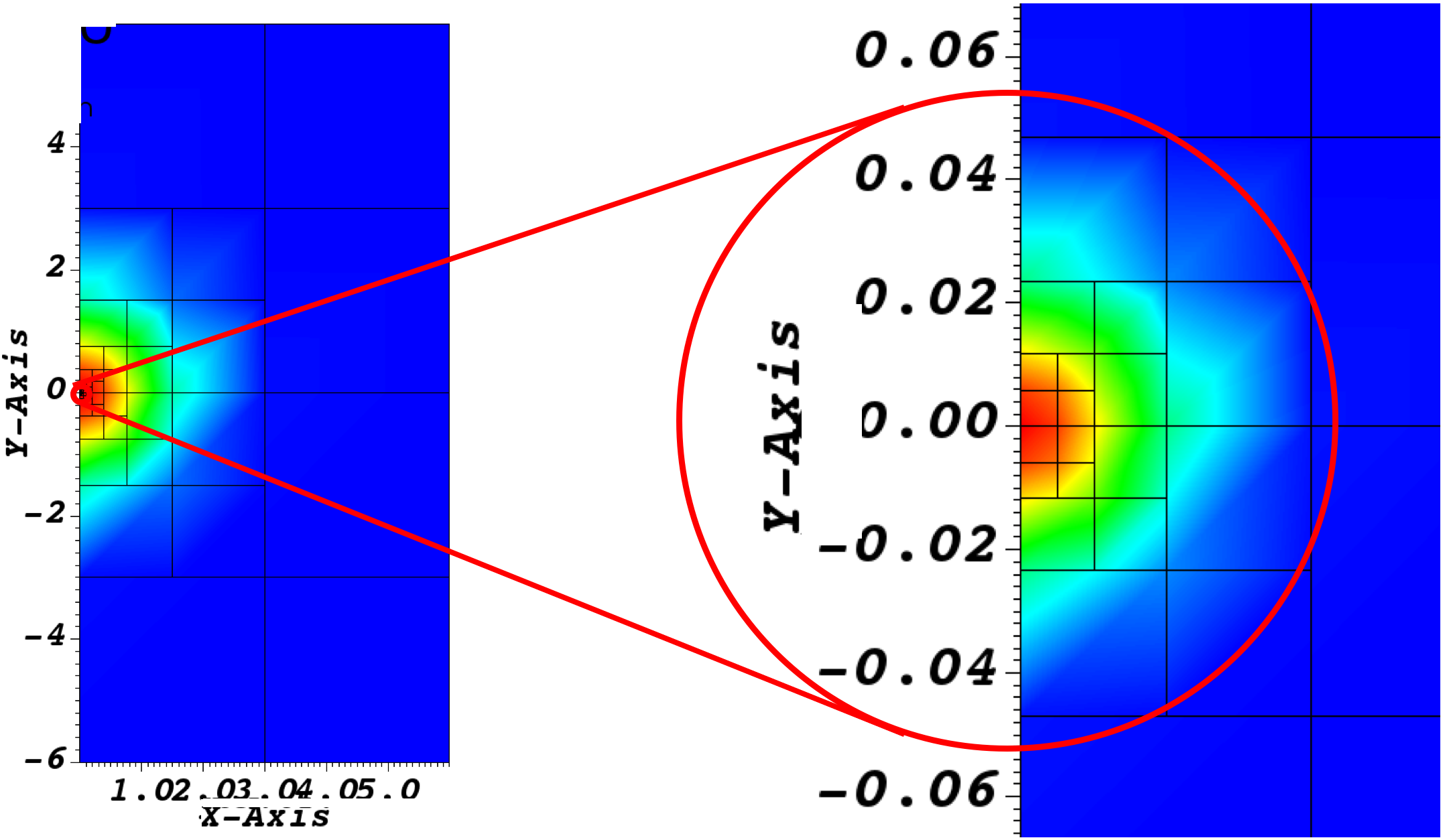}
\caption{Mesh for electron-deuterium plasma with Maxwellian distribution in units of electron thermal velocity. Electron distribution on global domian (left); detail with deuterium distribution (right). Visualization artifacts from linear interpolation in Visit}
\label{fig:grids}
\end{center}
\end{figure}

\subsection{Landau thermal quench code structure}
\label{sec:structure}

\begin{figure}[ht!]
\begin{center}
\includegraphics[width=1.01\linewidth]{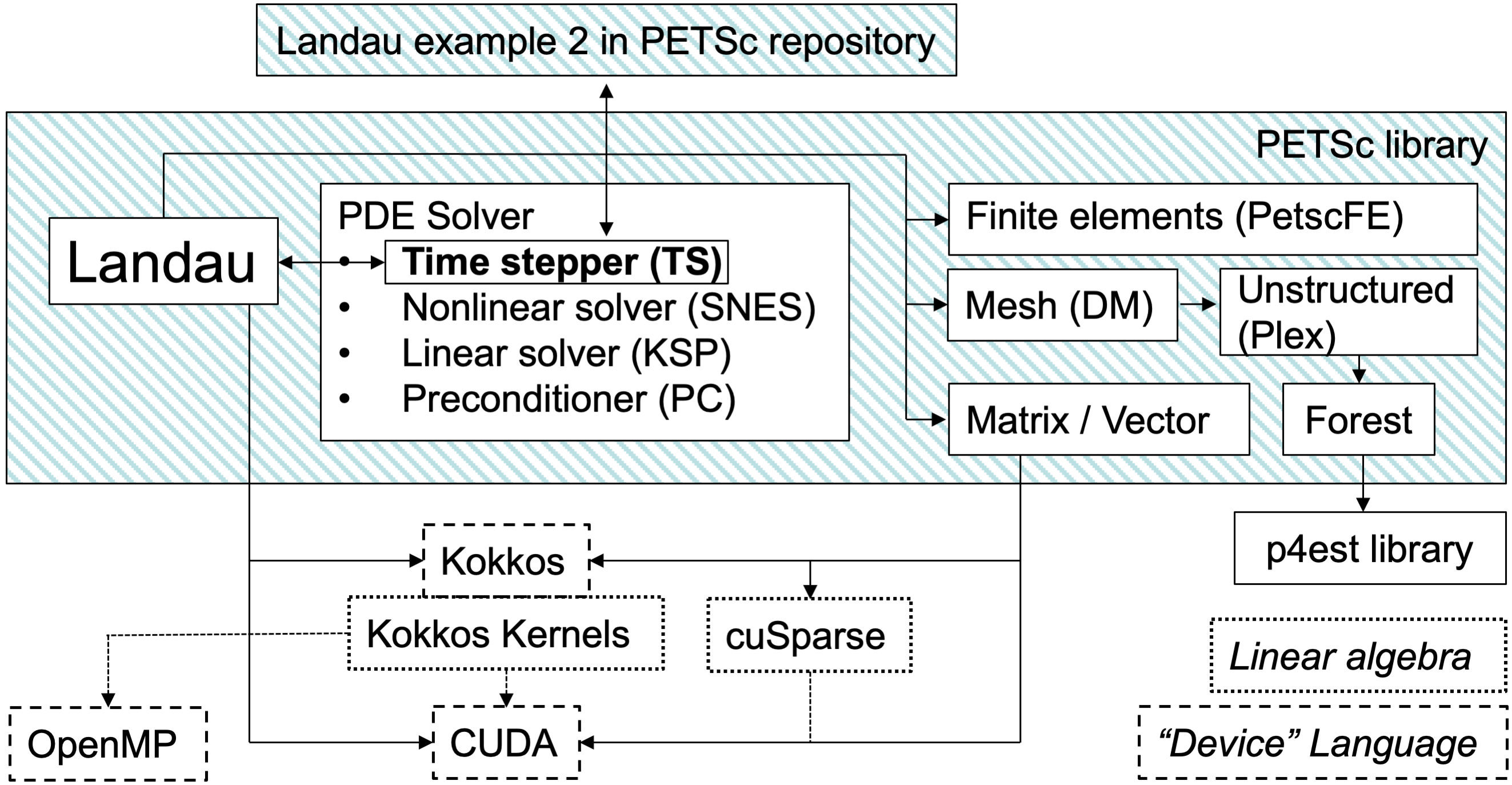}
\caption{Structure of the Landau thermal quench code as a PETSc example, the abbreviated component graph in the PETSc library, third party libraries and ``device" languages}
\label{fig:diag1}
\end{center}
\end{figure}

Figure \ref{fig:diag1} sketches the Landau thermal quench code structure.
PETSc is composed of a core PDE solver stack, discretization support (finite elements in this case), data (mesh) management, unstructured mesh management, adaptive mesh support and interfaces to device linear algebra packages that augment PETSc's build-in CPU linear algebra.
Not all connections are show here, such as Kokkos can be built with cuSparse or Kokkos Kernels on NVIDIA, the solver stack interfaces with the matrix and vector, and mesh classes.
HIP and SYCL back-ends are under development and mirror the CUDA back-end.
The Kokkos-HIP structure is not shown and it also mirrors the Kokkos-CUDA structure.
PETSc supports downloading and building third party libraries automatically during a configuration phase and integrating them with PETSc.
{\it P4est, Kokkos, Kokkos Kernels, cuSparse}, etc., are such libraries.

\subsection{CUDA and Kokkos implementations of Landau}
\label{sec:cudakok}

Two versions of this solver have been developed for the CUDA programming model, one written in CUDA using cuSparse,  the other in Kokkos using Kokkos Kernels \cite{CarterEdwards20143202}.
The CUDA and Kokkos versions of the kernel are similar but there are some differences.
The Kokkos syntax is higher level than CUDA.
In particular, Kokkos provides variable length arrays for the shared memory buffers, whereas the CUDA version uses array sizes fixed at compile time.
Kokkos provides a parallel reduction method, which supports reductions on general C++ objects that are equipped with obvious methods like a default constructor, a copy constructor and an {\it add} method.
The CUDA version parallelizes this inner integral manually (see Algorithm \ref{algo:1}).
Each thread accumulates a small vector and matrix for each species, and a warp shuffle efficiently broadcasts the sum of these partial integrals to all threads.
The Kokkos version hides this machinery in the parallel reduction method.
{Unlike CUDA, Kokkos is designed to be portable across vendors \cite{CarterEdwards20143202}.}


As far as the impact of multiple back-ends in PETSc, the vector interface for both back-ends is about 2,000 lines of code and the matrix interfaces are about 10,000 and 2,000 lines of code for the CUDA and Kokkos back-ends, respectively.
As far as the Landau code, the common CPU code, which includes a CPU implementation of the Landau kernel, is about 2,500 lines of code, and the each of the GPU back-ends is about 700 lines of code.
While maintaining three versions of the kernel (CPU, CUDA and Kokkos) imposes some overhead, the availability of several ``platforms" for development is useful to provide baseline performance of well optimized CUDA, and to allow incremental development from simple C code on the CPU, to Kokkos-CPU, to Kokkos-CUDA and finally to CUDA. 
For further details on performance portability in PETSc see Mills et al. \cite{mills2020performanceportable}.


\subsection{Algorithm for the CUDA programming model}
\label{sec:algo}

As developed in \cite{AdamsHirvijokiKnepleyBrownIsaacMills2017}, the element and integration point loops in the inner integral are merged and the data is packed into vectors for efficient processing ({ $r$, $z$, $w$, $f$ and $df$ in Algorithm \ref{algo:1}}).
Here, the data is transposed into a structure of arrays for GPU processing, from the arrays of structures used for vector architectures.
The outer loop over elements is parallelized in the CUDA programming model by putting one element in each ``league" member in Kokkos and the (x) dimension of the block grid in CUDA, and on one V100 SM or MI100 arithmetic unit (simply referred to as an SM herein) in either case.
The integration points are similarly parallelized into Kokkos thread teams and the (y) dimension of the CUDA thread block ({see Algorithm \ref{algo:1}}). 
The threads in Kokkos' ``ThreadVectorRange", and the (x) dimension of the CUDA thread block, compute the parallel reduction.
All threads on the SM participate in the finite element assembly.

\subsubsection{CUDA language optimizations}

The CUDA block size is chosen to be 256 or less threads. 
The second (y) dimension of the thread block is mapped to integration points; its size is dictated by the order of the elements.
The number of threads in the other (x) dimension is chosen to be a power of two, such that the total number of threads is less than or equal to 256.
Q3 elements (cubic finite element quadrilaterals) have 16 integration points, which corresponds to a block dimension of 16x16. 
Each SM processes one element.

Coalescing global memory access is important to maximize global memory throughput on the GPU. 
The 1D input arrays are stored in a structure of array format for this purpose. 
When accessing the 2D matrix of field values and derivatives, threads are mapped to the leading dimension of the matrix element to maximize coalesced access. 
The inner integral of (\ref{eq:Kalpha}), lines 3-11 in Algorithm \ref{algo:1}, is the most expensive part of the computation.
It is important to reduce redundant memory access and use fast memories as much as possible.
The $\beta$ terms of the integral are shared by all the integration points within an element.
All $\beta$ terms can be prefetched into shared memory.
The Landau tensors $\mathbf{U}^K$ and $\mathbf{U}^D$ can be pre-computed and stored in registers. 
The integration results can also be accumulated in registers. As a result, the inner integral loop accesses only registers and shared memory. The partial integral results stored in registers of different threads are accumulated into the final integral results using warp shuffle instructions. Finally, shared memory is used to store the accumulated $\mathbf{D}$ tensors and $\mathbf{K}$ vectors from the inner integral.

\subsection{GPU assembly of sparse matrices}
\label{sec:ass}

PETSc provides a compressed sparse row storage matrix with an object-oriented interface, written in C, where data is inserted with a ``MatSetValues" method that takes a dense 2D matrix of values and the global row and column indices to which the data is added.
Recently a GPU coordinate format (COO) matrix and a GPU version of the traditional interface in CUDA and Kokkos have been added for GPU assembly.
The Landau solver uses the traditional interface, which currently requires the matrix to be assembled once on the CPU.
Subsequent assemblies can then take place on the GPU.
The COO interface does not require this CPU assembly stage.
Both GPU assembly interfaces are works in progress.
The cost of the CPU step is amortized for the Landau solver because a transient analysis would use the metadata for many time steps.

GPU assembly requires that contention between elements running in shared memory be resolved.
Three basic approaches to this, in increasing order of code complexity, are atomic ``fetch and add", graph coloring to assemble several matrices in parallel, summing them when complete, and domain decomposition with some resolution process at the domain boundaries.
{Only the atomics approach has been released in PETSc}.


\subsection{Linear solver for multi-species Landau operator}
\label{sec:solver}

The implicit time integration for the advance of the collision operator requires an algebraic solver.
Direct solvers are attractive because of their low constants in complexity and the small sizes of these grids does not incur the cost of their sub-optimal asymptotic scaling.
Additionally, a {shared memory, or MPI serial,} direct solver can be written with only a few kernel launches relatively easily whereas fast iterative methods, like multigrid \cite{UTrottenberg_CWOosterlee_ASchueller_2000a}, for unstructured problems are more complex, although algebraic multigrid does work well mathematically on these (elliptic) problems \cite{ADAMS201935}.
PETSc relies on third party libraries, such as SuperLU and MUMPS \cite{superlu99,MUMPS:2}, for parallel direct GPU solvers.
 {The Landau matrices, however, are much smaller than the regimes that these solvers target and they did not perform well.
In response we wrote a custom CUDA LU factorization and solve for this project.}

{This CUDA band solver uses reverse Cuthill–McKee (RCM) ordering \cite{10.1145/800195.805928}, which naturally produced a block diagonal matrix in multi-species problems and is designed to minimize bandwidth.
Band sparse matrix storage stores the main diagonal and $UBW$ diagonals directly above the main diagonal and $LBW$ diagonals directly below the main diagonal.
Jacobians are generally structurally symmetric so that $B\equiv UBW=LBW$.
The standard outer product form of banded LU factorization is used (Algorithm 4.3.1 \cite{GoluVanl96} ).
This algorithm computes, for each row $i$, a $BxB$ outer product update of a dense sub-matrix with $A(i+1:,i)*A(i,i+1:)$.
}

Band solvers are attractive because of the simplicity of their kernels.
The band solver exploits the independent solves for each species and uses the group synchronization function in CUDA to allow for more than a single SM to process each species' matrix factorization.
Kokkos does not provide a group synchronization method and we have not implemented the band solver in Kokkos.


\subsection{Single grid vs single grid per species}
\label{sec:multi-grid}
 
 The Landau grids in this paper uses a single grid with a degree of freedom for each species.
 One can also use a grid for each species, which has the advantage that each grid can be scaled to resolve the distribution of each species efficiently.
 The multiple grid approach can be viewed as a simple type of mesh adaptivity and with it Cartesian grids can be used for efficiency \cite{Hager2016}.
 
There are advantages and disadvantages to the single grid vs the multiple grid approach.
To understand the complexity issues, consider a model where all species require some given mesh to resolve a Maxwellian distribution.
For instance, a typical one-species grid with 20 cells is shown in Figure \ref{fig:grids}, with a Maxwellian distribution and a typical domain size of five thermal velocity ($v_{th}$) units.
\begin{figure}[htbp]
\begin{center}
\includegraphics[width=1.\linewidth]{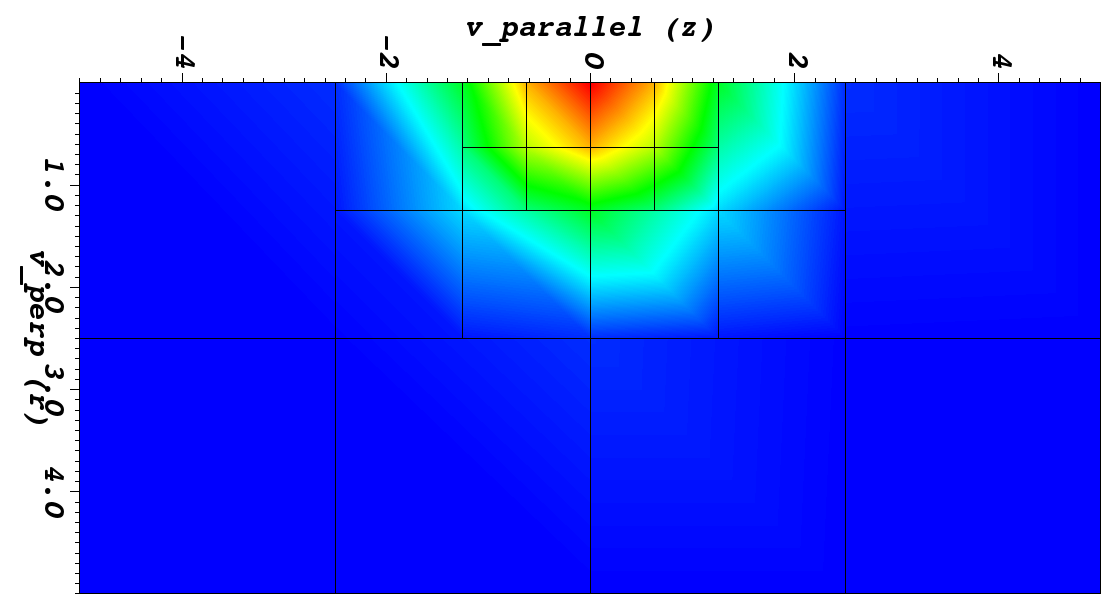}
\caption{Maxwellian with 20 cells and domain size $\mathbf 5v_{th}$ \\ (visualization artifacts from linear interpolation)}
\label{fig:grids}
\end{center}
\end{figure}
With Q3 elements there are 128 integration points in a radius of a bit over one thermal radii, which resolves the total energy of the Maxwellian with about five digits of accuracy.
An equivalent Cartesian would require 128 total cells, a $6.4x$ increase. 
This cost is a function of the desired accuracy.
High accuracy and large domain size benefit more from mesh adaptivity.

Consider a plasma with ten species, electrons, a light ion like deuterium and eight effective ionization states of a heavy ion like tungsten.
All with the same thermal temperature.
This case reflects, for instance, a plasma with impurities from the wall of a tokamak.
A single grid with electrons and tungsten requires about 74 cells to provide similar resolution as the single species grid with 20 cells.
The deuterium is highly resolved because its thermal velocity is bracketed by electron and tungsten thermal velocities.
All eight tungsten spices can share a grid because they share one thermal velocity.
The thermal velocities of electrons, deuterium and tungsten are well separated such that one 20-cell grid cannot resolve any two species.
Thus, this model requires three 20-cell grids or one 74-cell grid. 

Three quantities of interest in the complexity of the Landau operator are shown in Table \ref{table:1}: the number of integration points ($N$), the number of Landau tensor calculations ($IP\left(IP-1\right)/2$ if symmetry is exploited, which we do not), and the number of equations in the solve.
\begin{table}[ht]
\begin{center}
\begin{tabular}{c  r r r} 
\hline
\# grids    & $N$ integration points & \# Landau tensors ($N^2$) & $n$ \\  
\hline\hline
  1    & 1,184 &  1.4M   & 8,050 \\
  3    &    960 &  0.9M & 1,930\\
  10  & 3,200 &  10.2M & 1,930 \\ \hline
\end{tabular}
\caption{Cost for the Landau operator with 10 species vs number of grids: number of integration points $N$, number of Landau tensor calculations and number of equations $n$ }
\label{table:1}
\end{center}
\end{table}
The number of equations, or number non-constrained vertices in the (nonconforming) adaptive mesh, is taken from runs of the code.
The 20-cell grid generates 193 vertices and the 74-cell grid generates 805 vertices.
Q3 elements have 16 integration points in each element.
Clearly, multiple grids with multiple species per grid is optimal with much smaller number of equations to solve than the single grid approach and much fewer Landau tensor calculations than the 10-grid approach.

Generally speaking, low-resolution requirements allow for the use of fewer cells.
Large separation of thermal velocities and low-resolution requirements benefit from more grids because one grid cannot resolve the disparate velocity scales efficiently without excessive over resolution.
{Species with the similar thermal velocities (say within $2x$ or more) can, and should, share a grid.
Clusters of thermal velocities, in the spectrum of thermal velocities of a given problem, should share a grid if multiple grids are supported.
}


\section{Runaway electrons and thermal quench plasmas}
\label{sec:physics}

Effective collision operators are useful for understanding the physics governing dynamic and/or highly structured distribution functions, typically driven by sources and external forcing.  The effect of a fast thermal quench on a current carrying plasma has all of these characteristics, and is among the most important problems in plasma physics.  In a thermal quench, the thermal energy of the electrons is rapidly lost due to either a large source of cold electrons being introduced as a source, or by some other means such as parallel heat loss along open magnetic field lines, or some combination thereof.  For high temperature plasmas, the electron thermal energy can easily be lost in a time on the order of or less than a typical electron collision time.  Under these conditions, and with the introduction of cold impurities, possibly with multiple ion charge states, the resulting distribution can be far from Maxwellian and strongly time dynamic.

In particular, because the mean free path and average collision time increases with particle energy within the distribution, a fast thermal collapse can cool the bulk plasma to low energy, but leave behind the higher energy tail of the distribution which would need more time to equilibrate. This higher energy tail can become a seed population for further acceleration and growth given an electric field to accelerate it.  In plasmas with high current, such as toroidal magnetically confined plasmas, the resulting highly collisional low energy part of the distribution will generate a large electric field which can accelerate the higher energy seed to even higher energy.  As the collisionality of this seed reduces even further with its increasing energy, a runaway condition can occur, accelerating these electrons to GeV energies.  The generation of runaway electrons in tokamak plasmas is of great concern to fusion energy scientists \cite{boozer2017runaway}, but can also occur in natural conditions such as lightning and solar flares.


\subsection{Spitzer resistivity} 
\label{sec:spit}

A model for plasma resistivity is critical for both the thermal quench model and for verification of any collision operator.
A classic expression for plasma resistivity, known as Spitzer resistivity \cite{Spitzer1950}, is derived from a model similar to the FP-Landau model that is also diffusive and effective for small angle collision dominated plasmas.
This expression for the resistivity parallel to the electric field is given by 
\begin{equation}
\label{eq:spitzer}
\begin{split}
\eta = {\frac {4{\sqrt {2\pi }}}{3}}{\frac {Ze^{2}m_{e}^{1/2}\ln \Lambda}{\left(4\pi \varepsilon _0\right)^{2}\left(k_{\text{B}}T_{e}\right)^{3/2}}} F(Z),  \\ 
F(Z)={\frac {1+1.198Z+0.222Z^{2}}{1+2.966Z+0.753Z^{2}}},
\end{split}
\end{equation}
where $Z$ is the effective ionization of nuclei, $k_{\text{B}}$ is Boltzmann's constant and $T_e$ is the electron temperature in kelvins \cite{wiki:Spitzer}.

\subsection{Verification with Spitzer resistivity}
\label{sec:spitzer_a}


An equilibrium plasma with a small applied electric field $E_z$ develops a current that asymptotes to a quasi-equilibrium.
This current can be computed with the integral $J_z =\sum_{\alpha} \int \limits_{\Omega}d{x} 2\pi \vec{x}_{r} q_\alpha  \vec{x}_{z} f_\alpha({x})$, where $q_\alpha$ it the charge of species $\alpha$.
Computed resistivity is then defined as $\eta = E/J_z$.
It has been observed that this $\eta$ is not sensitive to (modest) electric field strength.
Plasma resistivity is a collisional phenomenon and the FP-Landau model should approximately converge to Spitzer resistivity (\ref{eq:spitzer}).
We observe that this FP-Landau code with a deuterium plasma converges to an effective plasma resistivity that is about $1\%$ lower than Spitzer resistivity (see Appendix \S\ref{sec:ad}).
This implies that the Spitzer model is in effect ``seeing" more collisions than the FP-Landau model.



\begin{figure}[htbp]
\begin{center}
\includegraphics[width=.9\linewidth]{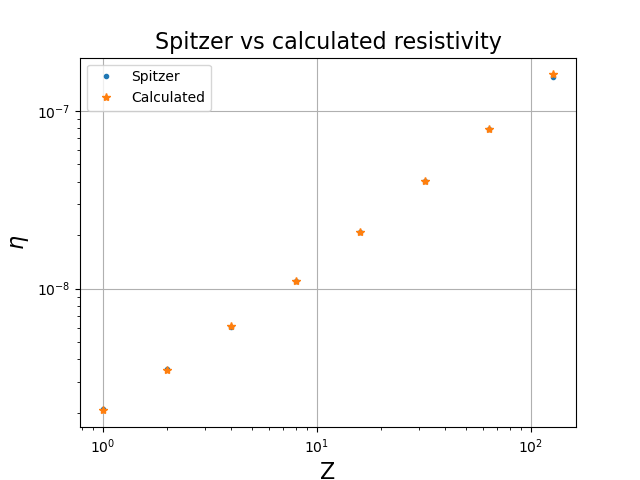}
\caption{Calculated $\eta = E/J$ and Spitzer $\eta_z$ as a function of $Z$}
\label{fig:z}
\end{center}
\end{figure}

As a qualitative verification test, Figure \ref{fig:z} plots the value of $\eta = E/J$ to the Spitzer $\eta$ as a function of the effective ionization $Z$. 
This data is with a 176-cell mesh of Q3 elements and the solver for the $Z=128$ case was not fully converged, which probably accounts for the noticeable discrepancy in this case.

\subsection{Vlasov-Maxwell-Landau thermal quench}
\label{sec:vmlq}

\begin{figure*}[ht]
\begin{center}
\includegraphics[width=1.\linewidth]{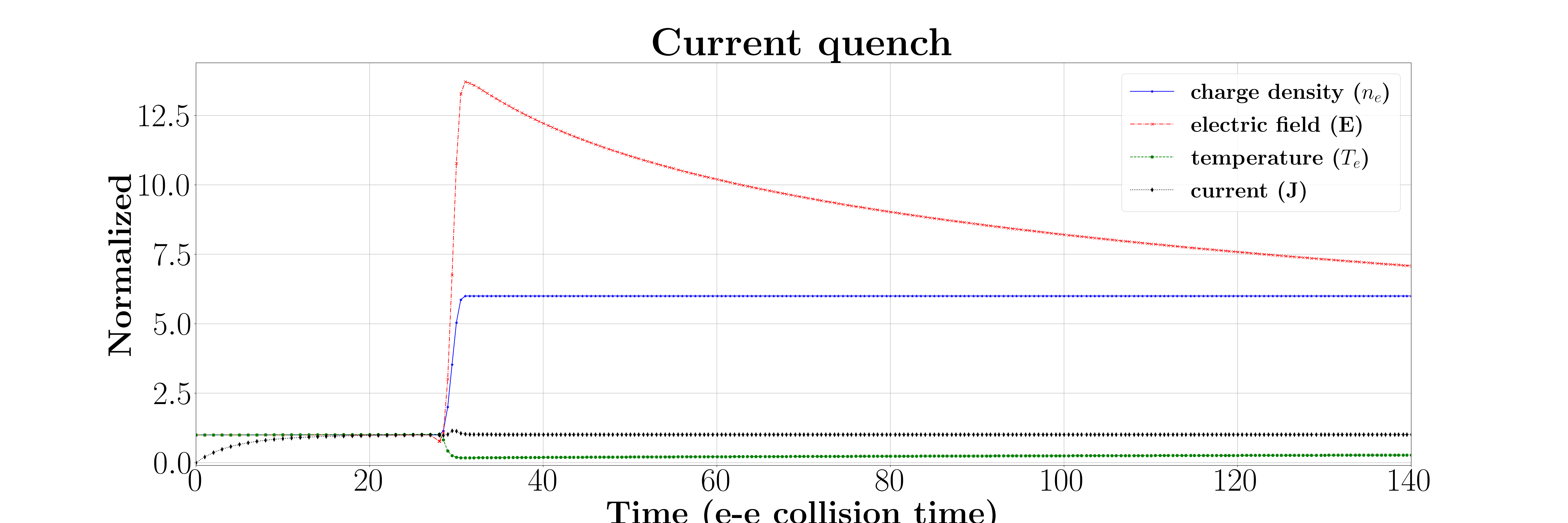}
\caption{Profiles of thermal quench model with cold plasma injection}
\label{fig:quench}
\end{center}
\end{figure*}

The thermal quench model begins like the Spitzer resistivity test (\S\ref{sec:spitzer_a}), but when a quasi-equilibrium current is detected it switches to computing $E \leftarrow \eta J$ with Spitzer $\eta$, leaving the plasma in a  quasi-equilibrium.
A pulse of cold ions is then injected with the source term in (\ref{eq:weak-form}), which lowers the temperature and thereby increases $\eta$ via Spitzer resistivity.
The electron temperature $T_e$ in (\ref{eq:spitzer}) is computed similarly to $J_z$ in \S\ref{sec:spitzer_a}.
This in turn increases $E$, which accelerates energetic electrons and increases the kinetic part of the total current $J$, but at a slower time scale.
The increase in $E$, combined with reduced friction on high energy electrons from mid-velocity electrons that have been depleted by the quench, can accelerate fast electrons even further.
As a fast electron population separates from the slow electrons they are subject to less friction and can continue to accelerate forming a population of seed runaway electrons.

Figure \ref{fig:quench} shows profiles of normalized charge density $n_e$, current $J$, electric field $E$, and electron temperature $T_e$, as a function of time in electron-electron collision time units, from a numerical experiment where the initial $E = 0.5 E_c$.
$E_c$ is the Conner-Hastie critical electric field strength for runaway electrons \cite{Connor1975,Dreicer59}.
The electron density is conserved exactly and thus with sufficiently accurate time integration the profile $n_e$ is the prescribed sinusoidal source function.
The total mass injected by the model is five times the initial density, which is observed to high accuracy.
The collapse of the temperature, and its gradual rise from electric field heating, is observed.
This test shows that this model is able to qualitatively produce the expected dynamics of a thermal quench, however this model is not complete enough to generate seed runaway electrons without an unrealistically high electric field.


\section{Performance experiments}
\label{sec:perf}

Kinetic applications commonly use operator split time integrators, where the simplectic Vlasov system and the metric collisions are alternately advanced.
Each configuration space point advances the collision operator -- independently -- which provides significant task parallelism in a real application.
An application would run thousands or more of these vertex solves in a collision advance step on each GPU.

To mimic an application, these experiments use one ``node" of a given machine with many MPI processes asynchronously launching jobs on the GPUs.
Running on a whole node applies pressure on the entire memory system as would occur in an application.
An MPI harness code (the Landau ex2.c example in PETSc) runs the same simulation on each MPI rank.
A flat MPI model provides asynchronous dispatch without any explicit asynchronous code.
NVIDIA's Multi-Process Service (MPS) system aids in scheduling the GPU with input from multiple streams from MPI processes.
In this context, the most important figure of merit is throughput: Newton iterations per second.
This metric factors out the specifics of the time integrator and non-linear solver tolerance, etc., which is application dependent.
Throughput is defined as the total number of Newton iterations times the number of instances of the problem run in parallel (MPI processes), divided by the simulation time after setup costs that are amortized in a long running simulation.

The test problem is similar to the deuterium plasma in \S\ref{sec:amr} and \S\ref{sec:vmlq}, but with an additional eight species of Tungsten with different ionization states, which is typical of a production run with impurities from the wall of the tokamak, and with 80 Q3 elements, run for 100 time steps.





\subsection{IBM POWER9 / NVIDIA V100 }
\label{ssec:perf-summit}

The CUDA and Kokkos-CUDA back-ends are tested with one Summit node: two IBM POWER9 processors with six NVIDIA V100 GPUs.
Each POWER9 has 21 cores (7 cores per GPU) and each core has four hardware threads.
MPS and CUDA-11 were not compatible on Summit at the time of this writing and thus the Summit results use CUDA-10;
These IBM experiments use CUDA v10.1 and gcc v6.4 with -O3 (see Appendix \S\ref{sec:ad} data and reproducibility description).

Tables \ref{tab:cuda10} and \ref{tab:kokkos10} report the throughput on one Summit node with the CUDA and Kokkos-CUDA back-ends, with respect to the number of cores per GPU and number of processes per core.

\begin{table}[h!]
\centering
\caption{CUDA-10, V100 Newton iterations / sec}
\label{tab:cuda10}
\begin{tabular}{lrrrrr}
\toprule
cores/GPU &     1 &     2 &     3 &     5 &     7 \\
process/core &       &       &       &       &       \\
\midrule
1            &   849 & 1,683 & 2,487 & 4,044 & 5,504 \\
2            & 1,102 & 2,142 & 3,177 & 5,094 & 6,838 \\
3            & 1,096 & 2,189 & 3,252 & 5,239 & 7,005 \\
\bottomrule
\end{tabular}
\end{table}

\begin{table}[h!]
\centering
\caption{Kokkos-CUDA-10, V100 iterations / sec}
\label{tab:kokkos10}
\begin{tabular}{lrrrrr}
\toprule
cores/GPU &     1 &     2 &     3 &     5 &     7 \\
process/core &       &       &       &       &       \\
\midrule
1            &   792 & 1,542 & 2,265 & 3,511 & 4,849 \\
2            &   996 & 1,974 & 2,904 & 4,641 & 6,013 \\
3            & 1,010 & 2,044 & 2,982 & 4,805 & 6,193 \\
\bottomrule
\end{tabular}
\end{table}


The fastest throughput for all back-ends use all seven cores per GPU and two or three hardware threads per core, with a modest but consistent gain in using a second and usually a third hardware thread.
This data shows that CUDA is about $15\%$ faster than Kokkos-CUDA.
Given that Kokkos is a portable language, this performance penalty is not unexpected nor unreasonable.

\subsubsection{{Hardware utilization of the V100}}
\label{sec:perf-hw}


The matrix construction is split into the assembly of the Jacobian and the assembly of a scaled mass matrix as dictated by the time integrator.
The finite element mass matrix is the identity in weak forms and is added to the Jacobian in all time integrators.
The mass matrix replaces all of Algorithm \ref{algo:1} with $\mathbf{C} \leftarrow Transform\&Assemble\left({w[gip]s}, \mathbf{0},\mathbf{0}, \mathbf{B}, \mathbf{0} \right)$, where $s$ is a shift determined by the time integrator.
The analysis of the hardware utilization in the GPU kernel is divided into the analysis of the Jacobian matrix and the mass matrix construction.


The NVIDIA Nsight Compute tool is used to gather all the hardware metrics with a single process.
To collect meaningful hardware metrics the hardware resource must be fully utilized, which requires a 320-cell version of the test problem used in \S\ref{sec:perf}.
About $8$\% of the total matrix construction time is from the mass and thus about $92$\% is in the Jacobian in these tests.

For V100, the DFMA peak is 7.8 TFlop/s and DRAM peak is 890 GB/s. So the arithmetic intensity (AI) roofline turning point is at 8.8.
The Jacobian kernel is primarily compute bound with an AI of 15.8 and the FP64 pipe utilization is measured to be 66.4\%. 
The kernel achieved ~4.15 TFlop/s. This is 53\% of the peak DFMA throughput on V100, which corresponds to the roofline percentage. The roofline percentage is lower than the FP64 pipe utilization because only 64\% of the FP64 instructions were DFMA instructions. The rest are DMUL and DADD.

\begin{table}[h!]
\centering
\caption{Roofline data for Jacobian and mass operator}
\label{tab:roofline}
\begin{tabular}{lrrr}
\toprule
 & AI &  \% roofline & Bottleneck (utilization) \\
\midrule
Jacobian & 15.8   &  53\% & FP64 pipe (66.4\%)  \\
Mass    &    1.8  &  17\% & L1 cache (27\%) \\
\bottomrule
\end{tabular}
\end{table}

The mass kernel has an lower AI of 1.8. 
This is expected because the mass kernel only performs finite element assembly and sparse matrix assembly, which is mainly memory copy operations with very little computation. Thus it looks like the Jacobian without the inner integral and with a simpler inner loop in the finite element assembly. 
The 17\% roofline percentage comes from the 17\% DRAM utilization. 
However, DRAM is not the leading bottleneck for this kernel. 
The kernel has a L1 hit rate of 77\%. 
As a result, most of the memory traffic is from L1. 
The L1 utilization is higher than the DRAM utilization at 27\%, but it is still low.
The mass kernel is L1 latency bound.  

The main reason for the low L1 utilization is load imbalance in memory traffic between different threads. 
Elements with constrained faces, from mesh adaptivity, interpolate each matrix value associated with a constrained degree of freedom to four degrees of freedom in the global matrix with the Q3 elements used here.
The elements in these meshes have $0-2$ constrained faces.
Such imbalance leads to a subset of threads in a warp accessing a partial cache line, which is inefficient on the GPU.
Furthermore, the imbalance caused some threads to exit early, which reduces the achieved occupancy.
Both of those effects lead to a low L1 utilization. 

\subsection{AMD EPYC  / MI100  with Kokkos-HIP}
\label{ssec:perf-amd}

This section present preliminary data from an AMD EPYC  / MI100 node.
Data from the Kokkos-HIP back-end is generated with one node of Spock: a 64-core AMD EPYC 7662 ``Rome'' CPU, with two hardware threads per physical core, and four AMD MI100 GPUs.
We use gcc-7.5.0 and rocm-4.1.0 for these experiments (Appendix \ref{sec:ad}).
Table \ref{tab:hip} reports the number of Newton iterations per second on four MI100 GPUs and up to eight cores per GPU, with one and two processes per core, with Kokkos-HIP.
\begin{table}[h!]
\centering
\caption{HIP, MI100 Newton iterations / sec}
\label{tab:hip}
\begin{tabular}{lrrrr}
\toprule
cores/GPU &   1 &   2 &   4 &   8 \\
process/core &     &     &     &     \\
\midrule
1            &  88 & 169 & 281 & 353 \\
2            & 154 & 272 & 341 & 241 \\
\bottomrule
\end{tabular}
\end{table}
This data shows a speedup of $4x$ with eight cores, with good initial speedup, however performance rolls over with 16 processes per GPU.



\subsection{A64FX with Kokkos-OpenMP}
\label{ssec:perf-fug}

This section experiments with one node of the Fugaku machine, with one Fujitsu A64FX processor, using up to 32 of the available 48 cores partitioned into 4-32 MPI processes.
The GNU compiler v8.2 is used, which corresponds to OpenMP v4.5 (-Ofast -march=armv8.2-a+sve -msve-vector-bits=512, Appendix \ref{sec:ad}).
Table \ref{tab:fug} shows the matrix construction and total simulation times of a 10-time step version of the model problem, as a function of the number of MPI processes and number of OpenMP threads per process, with the Kokkos-OpenMP back-end.
\begin{table}[h!]
\centering
\caption{Jacobian construction and total time (sec) on one Fugaku node and total solve time of the 32 core case (diagonal)}
\label{tab:fug}
\begin{tabular}{lrrrrr}
\toprule
\multicolumn{1}{l|}{\#processes} & threads/processes \quad 8 &  4 &  2 &  1 & Total \\
\midrule
4   &  (19.3) &  38.1  & 75.3   &  150 & 25.1\\
8   &         & (38.1) &        &      & 45.9 \\
16  &         &        & (75.5) &      & 87.0 \\
32  &         &        &        &  (150) & 169.4 \\
\bottomrule
\end{tabular}
\end{table}

This data shows excellent thread scaling in that times are inversely proportional to the number of threads with four processes (top row) and the throughput (\#processes/time $\approx 5$) is almost constant with 32 cores (diagonal).
The total time is not as ideal (right column, linear in \#processes would be perfect), indicating that the rest of the solver is not thread scaling perfectly.
This data, with 208 Jacobian matrix constructions, delivers a throughput of $39$ Newton iterations/second in the four process, eight threads per process case.

\subsection{Comparative performance}
\label{ssec:perf-parts}

Table \ref{tab:parts10} reports timings for the single process per GPU case from Tables \ref{tab:cuda10}, \ref{tab:kokkos10} and \ref{tab:hip} and the 4 process, 8 threads per process case in Table \ref{tab:fug}.
The maximum value measured by any process is reported.
The Fugaku data is normalized from a 10-time step test data.

The Landau matrix construction and the LU factorization and solve are the major components to the total cost.
The Landau matrix construction includes GPU kernel work and some meta-data processing on the CPU.
The CPU algebraic solver, PETSc's LU solver, is identical for all platforms.

\begin{table}[h!]
\centering
\caption{Component times -- V100/Power9, MI100/EPYC and Fugaku}
\label{tab:parts10}
\begin{tabular}{lrrrrr}
\toprule
Device &  Total &  Landau &  (Kernel) &  factor &  solve \\
\midrule
CUDA        &   14.3 &     3.3 &       2.9 &     8.4 &    0.8 \\
Kokkos-CUDA         &   15.4 &     4.1 &       3.2 &     8.7 &    0.8 \\
Kokkos-HIP &   23.1 &    10.9 &      10.2 &     5.9 &    0.5 \\
Fugaku (normalized) &  250.7 &   215.1 &     209.5 &    16.1 &    1.5 \\
\bottomrule
\end{tabular}
\end{table}

On Summit, about $20\%$ of the time is spent on the GPU (`Kernel') and this kernel time is about $80\%$ of the total matrix construction time (`Landau').
This explains why using more processes per GPU improved performance significantly on Summit, as seen in Tables \ref{tab:cuda10} and \ref{tab:kokkos10}.
The Spock GPU kernel is under-performing relative to Summit, which results in a higher percent of the Landau time being in the Kernel, and the EPYC processor is about $2x$ faster than the Power9 as reflected in the (CPU) factorization and solve times.
Fugaku is also under-performing relative to Summit, about $95\%$ of the run time is in the Landau kernel (\S\ref{ssec:perf-fug}).

\subsubsection{Comparative Spock performance}
\label{ssec:spock}

The Spock data shows that the kernel is under-performing relative to the V100.
The AMD MI100 GPU has a peak performance of up to 11.5 TFLOPS and the V100 has a peak of about 7.8 TFLOPS.
Normalizing the data with respect to theoretical peak, the Kokkos-CUDA Landau kernel time in Table \ref{tab:parts10} of about $3$ seconds is about $5x$ faster than the MI100.

There are a few potential sources of this under-performance.
This data was collected soon after the publication embargo was lifted on Spock and ROCm may have been under active development.
The MI100 has 120 compute units as compared to 80 on the V100 and thus needs more work to be fully occupied.
Unlike the V100, the MI100 does not have hardware support for double precision atomic-adds in global memory (\S II \cite{Stone21}), which is used in GPU finite element matrix assembly.
Stone et. al. developed algorithms to reduce the use of atomics in finite element residual calculations \cite{Stone21}, which is similar to finite element matrix assembly, and observed significantly more speedup on the MI100 than the V100 with their algorithms indicating that atomics are a significant source of MI100 under-performance relative to the V100.
Additionally, Multiple processes should be able to saturate the GPU, however Table \ref{tab:hip} shows throughput rolling over with 16 processes per MI100.
This indicates that the AMD equivalent to MPS is not functioning well.
Note, we have informally observed a throughput speedup, on a typical high throughput case in Table \ref{tab:cuda10}, of about $3x$ with the use of MPS.



\subsubsection{Comparative Fugaku performance}
Using Top500 data to normalize the nodes, a V100 is about $2x$ more powerful than an A64FX node/processor.
Scaling the Fugaku data up by $1.5$, only 32 of the 48 cores are used, the throughput, normalized with the V100, is about 117 iterations/second.
Comparing this to the peak of about 1,000 iterations/second per V100, from Table \ref{tab:kokkos10}, indicates that the A64FX is under-performing by about a factor of $8.5$. 
The A64FX has $8$ vector lanes, suggesting a lack of effective auto vectorization from the Kokkos v3.4 back-end and the GNU compilers.

\subsection{Performance summary}
\label{sec:perf_sum}

Table \ref{tab:perf_sum} summarizes the throughput  and the relative normalized performance analysis of the Landau kernel from the four cases investigated in \S\ref{ssec:perf-summit} -- \S\ref{ssec:perf-parts}.
\begin{table}[h!]
\centering
\caption{Throughput, Newton iterations (N/sec), \\ normalized performance relative to Summit / CUDA, \\ for each machine / language }
\label{tab:perf_sum}
\begin{tabular}{lrrr}
\toprule
Machine / language &  N/sec  &  \begin{tabular}{@{}c@{}} hardware \\  (GPUs + cores)\end{tabular}   & \begin{tabular}{@{}c@{}}kernel   \\  (\% CUDA)\end{tabular}    \\
\midrule
Summit / CUDA        &  7,005 &  6 V100 + 42 P9  & 100 \\
Summit / Kokkos-CUDA &  6,193 &  6 V100 + 42 P9  & 90 \\
Spock / Kokkos-HIP   &  353   &  4 MI100 + 32 EPYC  & 20 \\
Fugaku / Kokkos-OMP  &  39    &  NA  + 32 A64FX   & 12 \\
\bottomrule
\end{tabular}
\end{table}

\section{Conclusion}
\label{sec:conc}
This paper shows that the Landau collision operator can be practical for plasma physics applications with the effective utilization of GPUs.
The focus of this work has been on ameliorating the \Order{N^2} complexity of the Landau kernel, however end-to-end performance of a plasma thermal quench model demonstrates the potential practical use of this operator.
We have shown 66\% FP64 pipe utilization on the V100 with the CUDA back-end and have measured comparable overall performance with Kokkos-CUDA.

A ramification of this optimization is that the compute time of the high throughput runs of the entire collision advance is dominated by lower order complexity terms.
In particular, the linear solves and vector operations need attention.
Though a custom GPU LU solver is available in PETSc, it is no faster than the CPU solver reported here (see repository data, Appendix \ref{sec:ad}).
A custom GPU iterative solver is under development to address this problem.
The solver and vector operations would benefit from the batching of multiple spatial points, to augment or replace the existing asynchronous (MPI) thread dispatch, to reduce the number of kernel launches.
This batching of spatial vertices in the collision advances is also under development.

Other potential areas of future work includes integration with global plasma models to investigate runaway electron physics, adding support for multiple grids for groups of species with similar thermal velocities, as is done by Hager et al. for each species\cite{Hager2016}, and continuing to improve the entire solver stack in PETSc for GPUs.

Artifacts and reproducibility instructions are publicly available (see Appendix \ref{sec:ad}).

\section*{Acknowledgments}
The authors would like to thank  the PETSc team for making this work possible, and the assistance of Christian Trott with Kokkos.
This work was supported in part by the U.S. Department of Energy, Office of Science, Office of Advanced Scientific Computing Research and Office of Fusion Energy Sciences, Scientific Discovery through Advanced Computing (SciDAC) Program through the FASTMath Institute and the SCREAM and HBPS Partnership Projects under Contract No. DE-AC02-05CH11231 at Lawrence Berkeley National Laboratory.


\bibliographystyle{IEEEtran}
\bibliography{IEEEabrv,the}

\appendices

\section{Nondimensional variables}
\label{sec:nondim}

The Vlasov-Maxwell-Landau system is nondimensionalized with a thermal temperature of electrons $T_e$, a reference velocity $v_0 = \left ( 8kT_e/m_e\pi \right)^{\frac{1}{2}} $ and by defining a velocity coordinate ${ \vec{x} =  \vec{v} / v_0}$.
The distribution function variable is noramalized with $\tilde f_{\alpha} = f_{\alpha}v_0^3/n_0$ with a number density $n_0$ (eg, $10^{20}$ for a typical fusion plasma).
Nondimensionalize time, $\tilde t = t/t_0$,  with a reference time scale 
\begin{equation}
t_0 = \frac{8\pi m_0^2\varepsilon_0^2v_0^3}{e^4\ln\Lambda_{ee}n_0}, \quad \text{and define} \quad
\tilde { \vec{E}} = \frac{ t_0}{v_0} { \vec{E}}, \quad
\tilde \nu_{\alpha\beta} = \frac{t_0n_0}{v_0^3} \nu_{\alpha\beta}.
\end{equation}
Further, $d \vec{x} = v_0^{-3} d \vec{v}$, $\mathbf{U}( \vec{x},{\bar{x}}) = v_0\mathbf{U}(\ \vec{v},{\bar{v}})$  and $\frac{\partial}{\partial  \vec{x}} = v_0\frac{\partial}{\partial  \vec{v} }$.
Note, $\tilde \nu_{ee}=1$.
Any physical velocity space moment is given by $\int \ \vec{v}^n fd\ \vec{v}=n_0v_0^{n}\int  \vec{x}^nF d \vec{x}$. 
Rewriting the equation in these dimensionless coordinates results in
\begin{equation*}
\label{eq:landau2}
\begin{aligned}
&\frac{\partial \tilde f_{\alpha}}{\partial \tilde t}+  \frac{ e_\alpha}{m_{\alpha}} \tilde{ \vec{E}} \cdot \nabla \tilde f_{\alpha} = \\
&\sum_{\beta} \tilde \nu_{\alpha\beta}\frac{m_0}{m_{\alpha}}\nabla \cdot\int \limits_{\bar\Omega} d{\bar{x}}\;\mathbf{U}( \vec{x},{\bar{x}})\cdot\left(\frac{m_0}{m_{\alpha}} \tilde {\bar{f}}_{\beta} \nabla \tilde f_{\alpha} - \frac{m_0}{m_{\beta}} \tilde f_{\alpha} {\bar  \nabla} \tilde {\bar {f}}_{\beta}\right) + S_\alpha\left(t\right).
\end{aligned}
\end{equation*}
Observe that this nondimensionalized form does not change the equations, only the units.


 

\section{Artifact Description and reproducibility}
\label{sec:ad}

The entire time integrator and solver for the Landau operator is publicly available in the PETSc library (https://www.mcs.anl.gov/petsc {\tt git clone https://gitlab.com/petsc/petsc.git}
).
The thermal quench model used in these experiments is as an example in PETSc (ex2.c in the Landau tutorials).

PETSc output files with performance data and provenance information, the python scripts that generated most of tables, build instructions for each platform and reproducibility instructions and verification data can be found with {\tt git clone https://gitlab.com/markadams4/landau\_ipdps22}.
This repository also include data with a batched GPU LU solver and details of timing breakdown for all of the test cases that is not included in this report.

\end{document}